\documentclass[a4paper,twocolumn,amsmath]{revtex4-1}
\usepackage{soul}
\usepackage{ulem}

\usepackage{adjustbox}
\usepackage{color}
\usepackage{graphicx}
\usepackage{subfigure}
\usepackage{amsmath}
\usepackage{amsfonts}
\usepackage{citesort}
\usepackage{bm}

\newcommand{\comment}[1]{}

\newcommand{\eg}{{e.\,g.\ }}
\newcommand{\ie}{{i.\,e.\ }}

\newcommand{\Power}{{\mathcal{P}}}
\newcommand{\dirderiv}{\partial}
\newcommand{\mode}[2]{{#1}^{\text{(#2)}}}

\newcommand{\eps}{\varepsilon}
\newcommand{\imag}{\mathrm{i}}
\newcommand{\total}{\mathrm{d}}

\renewcommand{\emph}[1]{\textit{#1}}

\begin{document}

\title{Stimulated plasmon polariton scattering}
\author{C.~Wolff }
\email{cwo@mci.sdu.dk}
\affiliation{Center for Nano Optics, University of Southern Denmark, 
Campusvej 55, DK-5230 Odense M, Denmark}

\author{N.~A. Mortensen }
\affiliation{Center for Nano Optics, University of Southern Denmark, 
Campusvej 55, DK-5230 Odense M, Denmark}
\affiliation{Danish Institute for Advanced Study, University of Southern 
Denmark, Campusvej 55, DK-5230 Odense M, Denmark}
\affiliation{Center for Nanostructured Graphene, Technical University of Denmark, DK-2800 Kongens~Lyngby, Denmark}

\date{\today}

\begin{abstract}
  Plasmon and phonon polaritons of two-dimensional (2d) and van-der-Waals 
  materials have recently gained substantial interest.
  Unfortunately, they are notoriously hard to observe in linear response 
  because of their strong confinement, low frequency and longitudinal mode 
  symmetry.
  Here, we propose an approach of harnessing nonlinear 
  resonant scattering that we call stimulated plasmon polariton scattering
  (SPPS) in analogy to the opto-acoustic stimulated Brillouin scattering (SBS).
  We show that SPPS allows to excite, amplify and detect 2d plasmon and phonon 
  polaritons all across the THz-range while requiring only optical components 
  in the near-IR or visible range.
  We present a coupled-mode theory framework for SPPS and based on this find 
  that SPPS power gains exceed the very top gains observed in on-chip SBS by 
  at least an order of magnitude.
  This opens exciting possibilities to fundamental studies of 2d materials and
  will help closing the THz gap in spectrocopy and information technology.
\end{abstract}

\maketitle

%%%%%%%%%%%%%%%%%%%%%%%%%%%%%%%%%%%%%%%%%%%%%%%%%%%%%%%%%%%%%%%%%%%%%%%%%%%%%%%%
\section*{Introduction} 
%\TODO{Length: 501 words; target: 500}

The study of plasmon polariton excitations in two-dimensional (2d) 
materials~\cite{Low:2017}
and the related class of van-der-Waals (vdW) materials~\cite{Basov:2016} 
has recently gained considerable attention since they provide the means to very 
tightly confine, guide and manipulate electro-magnetic fields from the few-THz 
range all the way into the mid-infrared.
Furthermore and unlike conventional plasmonics based on metals, 2d material
plasmon polaritons are highly sensitive to their electromagnetic, electronic
and chemical environment, allowing for great tuning flexibility as well as
suggesting their versatile use for sensing~\cite{Garcia-de-Abajo:2014,Xiao:2016,Goncalves:2016}. As an example, the extreme spatial light confinement paves the way to applications as varied as mid-infrared vibrational fingerprints of proteins~\cite{Rodrigo:2015}, control of symmetry-forbidden atomic transitions~\cite{Rivera:2016}, or single-photon nonlinear optics~\cite{Gullans:2013}.
However, their greatest strengths --- unrivaled mode confinement and operation
in a hitherto poorly explored frequency regime --- along with their peculiar 
mode symmetry also turn out to be a nuisance for experimental work.
To date, the most promising avenue to overcome the extreme wave number mismatch
between polaritons and free-space radiation is by scattering at discontinuities,
\eg introduced by the probe of a scanning near-field optical microscope (SNOM).
  or material discontinuities designed into the device~\cite{Autore:2019}.
  As an alternative approach, the generation of graphene plasmon polaritons 
  by difference frequency generation based on the intrinsic nonlinearities
  has been studied both theoretically~\cite{Yao:2014,Rostami:2017}
  and experimentally~\cite{Constant:2015}.
This has led to significant insights~\cite{Fei:2017}, 
but is somewhat inefficient and does not seem very practical for applications 
beyond fundamental research.
  We suggest that this can be further drastically enhanced and harnessed for 
  practical applications by borrowing ideas from the seemingly unrelated field 
  of opto-mechanics, specifically of Brillouin scattering.

% 222 words

Stimulated Brillouin scattering (SBS) is the inelastic, resonant and 
self-amplifying back-scattering of light from a propagating acoustic wave in 
matter~\cite{Boyd:2003,Boyd:1990,Eggleton:2013}.
From its initial status as an academic curiosity, it has soon proven invaluable 
to characterize mechanical properties of materials at GHz frequencies -- a 
difficult range for direct mechanical measuring techniques.
More recently, it has attracted considerable attention for the realisation
of flexible yet highly selective optical filters~\cite{Choudhary:2017}, novel 
light sources~\cite{Otterstrom:2018}, the processing and buffering of optical 
signals~\cite{Shin:2015,Stiller:2018} and for the generation and amplification of coherent 
hypersonic waves \eg following the concept of the so-called phonon 
laser~\cite{Vahala:2009}.
% 85 words

\begin{figure*}
  \includegraphics[width=\textwidth]{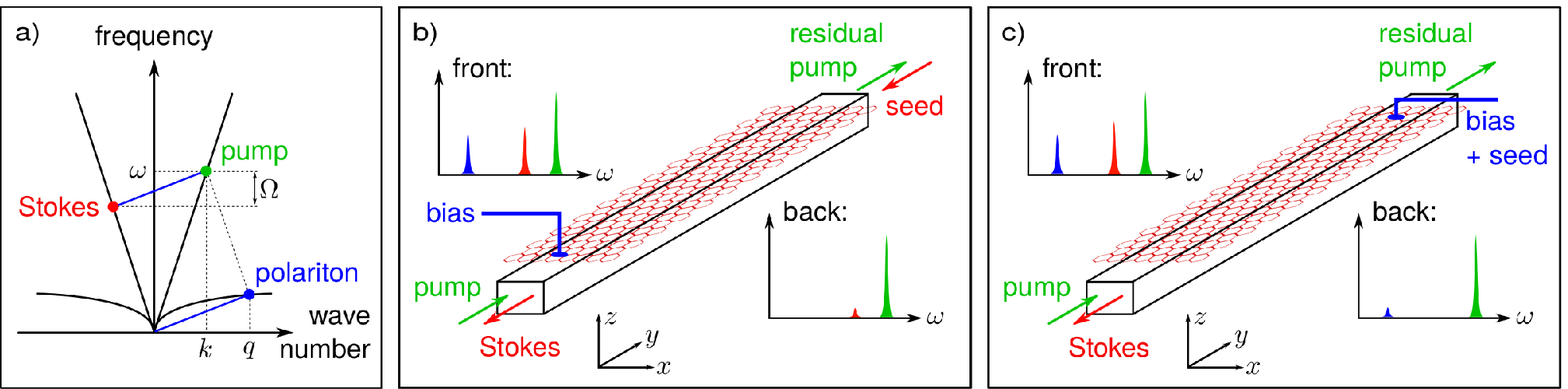}
  \caption{
    Illustration of the phase-matching in SPPS and schematics of two potential 
    realisations. \\
    (a) SPPS-interaction in the optical (straight solid) and polaritonic 
    (curved solid) dispersion relation: 
    The pump (green point, frequency $\omega$ and wave number $k$) is 
    scattered into the counter-propagating Stokes mode
    (red point) where the difference in frequency and momentum (blue line)
    matches the polaritonic dispersion relation.
    This polariton mode (blue point, wave number $q\approx2k$ for angular 
    frequency $\Omega \ll \omega$) is amplified along the waveguide.
    \\
    (b)
    Conceptual schematic of an SPPS-experiment in Stokes-seed configuration.
    A weak Stokes seed (red arrows) is injected at the rear end and amplified
    as is propagates towards the front end of a graphene-covered waveguide.
    This action is also illustrated by qualitative spectra where the heights
    of the colored peaks indicate to the relative amplitudes of the matching
    signals at the front and back of the waveguide.
    The bias voltage contact and the waveguide geometry shown here were 
    intentionally kept oversimplified at this stage and in reality would 
    require careful engineering.
    This configuration is best suited for the excitation of polaritons and
    for characterizing their dispersion relation, which can be tuned via the
    bias voltage contact.
    It is also the natural candidate for a first demonstration of SPPS as all
    input and output signals are in the optical domain and no injection or 
    detection of THz waves is required.
    \\
    (c)
    Conceptual schematic of an SPPS-experiment in polariton-seed configuration.
    A weak THz signal is injected through the bias contact (blue) and 
    amplified along the waveguide.
    Simultaneously, an optical Stokes is generated (illustrated by qualitative 
    spectra similar to panel b).
    This configuration is compelling because of the prospect to amplify or 
    optically detect weak THz signals.
% 201 words
  }
  \label{fig:principle}
\end{figure*}

We introduce the concept of inelastic, resonant and self-amplified 
scattering of light off propagating polaritons especially in 2d and vdW 
materials; a process that we refer to as stimulated plasmon polariton 
scattering (SPPS).
In analogy to the conceptually similar SBS, this will not only allow for 
the detection of 2d plasmons at the most convenient optical wavelength, 
but also to accurately measure both the frequency and damping at the given 
wave number.
In cases where the dispersion relation depends on a parameter (\eg the Fermi 
energy $E_F$), the wide-band nature of SPPS allows to characterize this 
dependency over 1--2 orders of magnitude. 
In the case of graphene plasmon polaritons (GPPs), this means that all regimes 
of the dispersion relation (pure intra-band scattering, inter-band corrections 
and potentially even nonlocal 
effects~\cite{Woessner:2014,Lundeberg:2017,Ni:2018,Dias:2018}) can be 
experimentally characterized in one setup.
Beyond this use in fundamental science, one can expect to harness SPPS for 
use in optical components, but especially for the tunable narrow-band 
amplification and optical detection of signals in the THz range. 
In the remainder of this paper, we describe the principle of SPPS and conclude
that it is experimentally observable in a standard silicon slot-waveguide 
covered with an appropriately biased graphene-monolayer.
% 194 words

  The proposed mechanism bears similarities with the excitation of graphene
  plasmons~\cite{Yao:2014} through difference-frequency generation (DFG) via 
  the intrinsic Pockels nonlinearity of graphene~\cite{Cheng:2017}, but our 
  aim is different.
  Our focus is not the intrinsic graphene nonlinearity that unlocks DFG, but
  rather the consequences of this nonlinearity on the interplay between 
  optical and polaritonic fields that are confined in a common long waveguide.
  We show that this will cause exponential gain for the THz polaritonic wave
  on a length scale well beyond its linear propagation length provided the
  nonlinear interaction is sufficiently strong.
  We find that among the nonlinear processes that can be found or introduced 
  in a system composed of a dielectric waveguide combined with a graphene 
  sheet, the ponderomotive nonlinearity~\cite{Wolff:2019} is sufficiently 
  strong, but we stress that it is not necessarily the only possibility.

%%%%%%%%%%%%%%%%%%%%%%%%%%%%%%%%%%%%%%%%%%%%%%%%%%%%%%%%%%%%%%%%%%%%%%%%%%%%%%%%
\section*{Results}

\subsection{Principle}
%\TODO{Length: 749 words; target: 1000}

The SPPS process is conceptually related to the well-understood SBS process in 
nano-scale waveguides:
in both cases a propagating low-frequency excitation with wave number $q$ 
modifies the local permittivity of a waveguide through a nonlinear process.
In SBS, the excitation is a sound wave and the nonlinear process is due to 
photoelasticity and the motion of the dielectric interface.
In SPPS, the excitation is a localized polariton and the nonlinear interaction 
is either due to an intrinsic nonlinearity of the polariton system or to a 
Pockels nonlinearity in the waveguide.
Thus, the polariton creates a travelling low-contrast grating in the 
waveguide, which can scatter an optical pump wave (with wave number $k$) into 
a counter-propagating Stokes wave if the difference in optical frequency 
and momentum matches the polaritonic dispersion relation (phase-matching).
Assuming that the Stokes shift $\Omega$ is small compared to the optical pump 
frequency $\omega$, the phase matching condition is approximately given by 
the ratio $q \approx 2k$ (see Fig.~\ref{fig:principle}).
In analogy to SBS, overall conservation of energy and momentum require that 
both the polariton field as the optical Stokes wave grow approximately 
exponentially along the waveguide.
This process exists as soon as there is an efficient nonlinear coupling
mechanism between the optical pump and the polariton.
Naturally, this nonlinear coupling must be strong enough that the process is 
not immediately quenched by the linear losses experienced by the plasmon 
polariton.
At least for THz graphene plasmon polaritons this is actually not as bad as
it might perhaps sound.
  Their propagation loss is in fact very comparable to that of GHz-range 
  acoustic phonons found in SBS as illustrated by the fact that both THz 
  graphene plasmons as well as GHz sound waves in technologically relevant 
  materials~\cite{Boyd:2003,Pant:2011,Eggleton:2013,Vanlaer:2015a,Vanlaer:2015b}
  have quality factors of orders of magnitude $10^1$--$10^3$.
Finally, neither SBS nor SPPS require long-range propagation of their 
respective non-optical excitations.
% 201 words

\begin{figure*}
  \includegraphics[width=\textwidth]{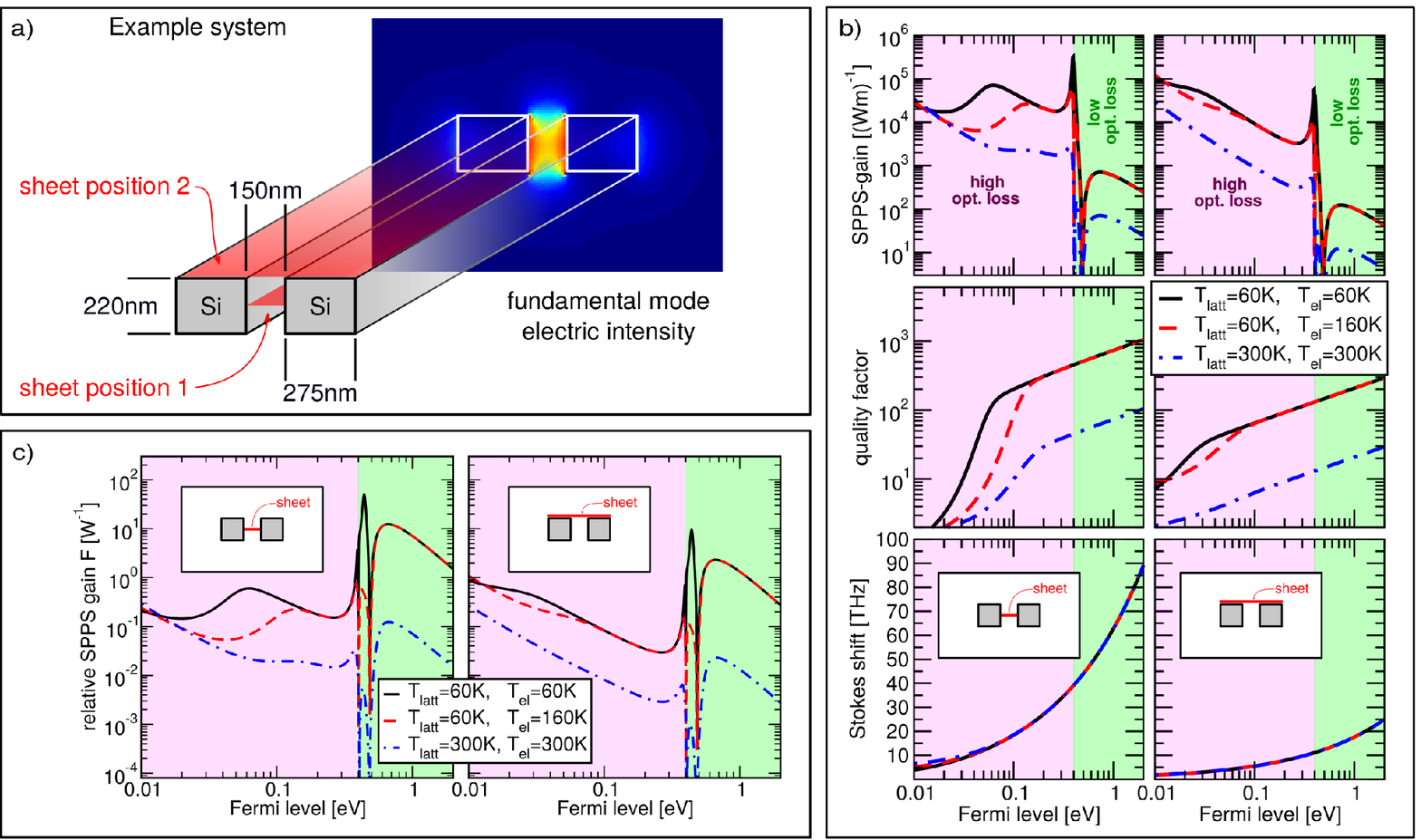}
  \caption{
    (a) Sketch of the example waveguide and electric
    intensity distribution of the fundamental mode at $1550\,\text{nm}$.
    The red planes indicate the two considered placements of the graphene
    sheet.
    \\
    (b) Total SPPS power gain, quality factor (\ie polaritonic loss)
    and Stokes shift (\ie polaritonic frequency) of the two
    example structures and at three different combinations of temperatures
    as functions of the Fermi level $E_F$.
    The temperatures are:
    $T=300\,\text{K}$, thermal equilibrium between phonon and electron gases
    (blue dash-dotted);
    $T=60\,\text{K}$, thermal equilibrium (black solid); 
    red dashed: lattice at 
    $T_\text{latt}=60\,\text{K}$ and electron gas at 
    $T_\text{el}=160\,\text{K}$ (red dashed) as a result of high pump powers
    %(see discussion in main text and supplemental material).
    (see discussion in main text and in Supplementary Note 2).
    The pink and green shaded areas indicate whether the optical pump can or 
    cannot drive inter-band transitions, leading to a high-loss and a low-loss 
    regime for the pump and Stokes signals.
    At the transition, the resonant inter-band contribution to the 
    ponderomotive force creates a pronounced peak in the gain.
    \\
    (c) 
    Relative SPPS gain figure $F = G / \kappa_a$ as a function of the Fermi 
    level.
  }
  % 102 words
  \label{fig:example_system}
\end{figure*}

\subsection{Theory}
We now introduce the basic theoretical framework.
In analogy to the theory of SBS~\cite{Wolff:2015a}, we describe the dynamics of
the three participating waves within the framework of coupled-mode theory, 
where we assume a strict slowly varying envelope approximation.
This means that the optical pump amplitude $\mode{a}{p}(y,t)$, the optical 
Stokes amplitude $\mode{a}{s}(y,t)$ and the plasmon polariton amplitude 
$b(y,t)$ all are assumed to vary slowly on the time scale of the slowest 
carrier in the system: the polariton frequency.
This all leads to the nonlinearly coupled equations
% 82 words
\begin{subequations}
\begin{align}
  & \dirderiv_y \mode{a}{p} + v_a^{-1} \partial_t \mode{a}{p} 
  + \kappa_a \mode{a}{p}
  = 
  - \imag \omega Q \Power_a^{-1} \mode{a}{s} b^\ast,
  \label{eqn:cme_pump}
  \\
  & \dirderiv_y \mode{a}{s} - v_a^{-1} \partial_t \mode{a}{s}
  - \kappa_a \mode{a}{s}
  = 
  - \imag \omega Q^\ast \Power_a^{-1} \mode{a}{p} b,
  \label{eqn:cme_stokes}
  \\
  & \dirderiv_y b + v_b^{-1} \partial_t b + \kappa_b b 
  = 
  - \imag \Omega Q \Power_b^{-1} [\mode{a}{p}]^\ast \mode{a}{s},
  \label{eqn:cme_gpp}
\end{align}
\end{subequations}
where we assume the waveguide to extend along the $y$-direction, $v_a$ is the
group velocity of the optical mode and $\Power_a$ is the power to which it 
has been normalized, and $\kappa_a$ is the optical decay parameter.
Their plasmonic counterparts are $v_b$, $\Power_b$, and $\kappa_b$, 
respectively.
Natural choices for the normalization powers are 
$\Power_a = \hbar \omega v_a / L$ and $\Power_b = \hbar \Omega v_b / L$, 
respectively, with the unit length of waveguide $L = 1\,\text{m}$.
The interaction is mediated by the nonlinear overlap between the optical
pump and Stokes modes and the permittivity change caused by the polaritonic 
mode:
% 79 words
\begin{align}
  \nonumber
  Q = & \langle \mode{\vec e}{p} | \ \Delta\eps(\mode{\vec e}{pol}) \ 
  | \mode{\vec E}{s} \rangle
  \\
  = & 
  \int \total^2 A \ \eps_0 \sum_{ijk} [\mode{e}{p}_i]^\ast \mode{e}{s}_j 
  \mode{e}{pol}_k \Delta\eps_{ij}(\mode{\vec e}{pol}),
  \label{eqn:coupling}
\end{align}
where the integral is carried out over the cross sectional plane of the 
waveguide, $\mode{\vec e}{pol}$ is the electric field distribution of the 
polariton mode and 
$\Delta\eps_{ij}(\mode{\vec e}{pol})$ 
includes all suitable second-order nonlinearities in the system.
This may include conventional Pockels nonlinearities $\chi^{(2)}$ in the 
waveguide or the polaritonic material or effects such as the ponderomotive
nonlinearity, which we base our example on.
In analogy to the theory of SBS~\cite{Wolff:2015a} we can easily derive from 
Eqs.~(\ref{eqn:cme_pump}--\ref{eqn:cme_gpp}) the stationary power gain
% 54 words
\begin{align}
  G = & \frac{2 \omega \Omega |Q|^2}{\Power_a^2 \Power_b \kappa_b},
  \label{eqn:gain}
\end{align}
which is the most interesting quantity in experiments.
  Besides inconsequential normalization constants and the 
  frequencies $\omega$, $\Omega$, its main ingredients are the nonlinear 
  coupling integral $Q$ and the polaritonic loss parameter $\kappa_b$.
  Therefore the reader concerned with loss in graphene might interpret $G$ as 
  a measure for how strong the nonlinear coupling is compared to the linear 
  polaritonic loss.
  This should be kept in mind when comparing the gain in successful SBS 
  experiments to the numerical values that we calculate further below.
The derivation of these equations and those underlying our numerical example
is beyond the scope of the main text and provided in Supplementary Note 1.
% 28 words

\subsection{Feasibility} 
We will now apply this theoretical framework to an illustrative waveguide 
geometry that has not been optimized for the strongest possible gain, but 
rather selected for its simplicity.
The main purpose of this example is to demonstrate that the concept is 
feasible, \ie that sufficient gain can be achieved in a realistic setting.
We focus entirely on the interplay of the opto-plasmonic nonlinearity and 
linear losses and leave important technological aspects of an actual 
experiment out.
This includes especially the exact method of doping, which can be achieved 
\eg electrostatically via a metal contact few microns above the structure, 
with direct electrical contacts~\cite{Ding:2015} or by chemical doping.
As a waveguide, we select a slot waveguide composed of two silicon beams with
rectangular cross section (dimension $220\times275\,\text{nm}^2$) separated by a 
$150\,\text{nm}$ air gap and operated at the standard telecom wavelength 
$1550\,\text{nm}$ corresponding to a photon energy of $0.8\,\text{eV}$.
We computed the electric field distribution of the fundamental mode (effective
index $1.37$) using the commercial finite element software COMSOL
(see Fig.~\ref{fig:example_system}a).
The slot waveguide was selected, because this configuration is known to support
waveguide modes that are mostly confined in the gap between the silicon beams
and are therefore an excellent starting point for nonlinear photonics in 
the silicon nanophotonics platform~\cite{Koos:2009}.
As a polaritonic system, we select a graphene monolayer that has been sandwiched
between thin layers of hexagonal boron nitride to isolate is from adverse 
substrate effects.
The polaritonic damping parameter $\kappa_b$ depends on $E_F$.
We calculate it from the Drude relaxation rates ($3.5\,\text{meV}$ at 300\,K 
and $0.35\text{meV}$ at 60\,K) that were experimentally 
determined for this type of ``vdW-sandwich''~\cite{Ni:2018} 
and plot it implicitly
in Fig.~\ref{fig:example_system}b as the polaritonic quality factor 
$\Omega / (2 v_b \kappa_b)$.
As mentioned before, this quality factor is the appropriate measure when
comparing to loss found in SBS and other related scattering processes.
% 143 words

We consider two possible placements of the graphene: either as a narrow 
ribbon inside the waveguide gap for maximal field enhancement or placed on top 
of the waveguide as a more practical arrangement.
  We note that the dispersion relation of such ribbons differs from that of
  infinitely extended graphene due to the lateral
  confinement~\cite{Goncalves:2020}.
In our example system, the second order nonlinearity stems entirely from the 
intrinsic ponderomotive force of the electron system in graphene. 
It is a direct consequence of the dependence of the optical properties 
on the Fermi level and has been studied based on a pure intra-band model for 
the optical conductivity~\cite{Mikhailov:2011}. 
In addition, we also consider the inter-band contributions and 
finite-temperature corrections to the ponderomotive force that 
we have published elsewhere~\cite{Wolff:2019}.
We should stress that while the ponderomotive interaction constitutes
a second-order nonlinearity, it must not be confused with the second-order 
nonlinear electric susceptibility $\chi^{(2)}$.
The latter is a third-rank tensor relating three electric field vectors,
and the former is the sensitivity of the conductance tensor to variations in 
the scalar carrier density and therefore in general described by only a 
second-rank tensor.

In Fig~\ref{fig:example_system}b, we show the calculated SPPS-gain, quality factor
and Stokes shifts of our example system for either sheet placement (left column:
inside the gap, right column: on top of the waveguide) and at a moderately
low temperature of $60\,\text{K}$ as well as at room temperature.
We find Stokes shifts throughout the entire THz-range, and for low 
Fermi-levels we predict SPPS-gains in excess of 
$10^4\,\text{(Wm)}^{-1}$ over a fairly large parameter range with peaks
approaching $10^6\,\text{(Wm)}^{-1}$.
This has to be compared
to $1-10\,\text{(Wm)}^{-1}$ for SBS in optical 
fibres~\cite{Nikles:1997,Kang:2009},
$100-1000\,\text{(Wm)}^{-1}$ in chalcogenide rib 
waveguides~\cite{Pant:2011,Eggleton:2013}
and $1000-10000\,\text{(Wm)}^{-1}$ in well-engineered silicon 
waveguides~\cite{Vanlaer:2015a,Vanlaer:2015b,Kittlaus:2017}.
This demonstrates that SPPS can be expected to provide levels of gain that are
very competitive with similar nonlinear processes such as SBS even though the 
graphene sheet introduces linear loss, which can limit the useful waveguide 
length to below $1\,\text{mm}$. 
To avoid confusion, we note that SBS-gains for bulk materials are usually 
specified in alternate units of m/W and can only be compared with our number 
when divided by a suitable mode area. 
% 241 words

%%%%%%%%%%%%%%%%%%%%%%%%%%%%%%%%%%%%%%%%%%%%%%%%%%%%%%%%%%%%%%%%%%%%%%%%%%%%%%%%
\section*{Discussion}
%\TODO{Length: 1243 words; target: 1000}

\begin{figure*}
  \includegraphics[width=\textwidth]{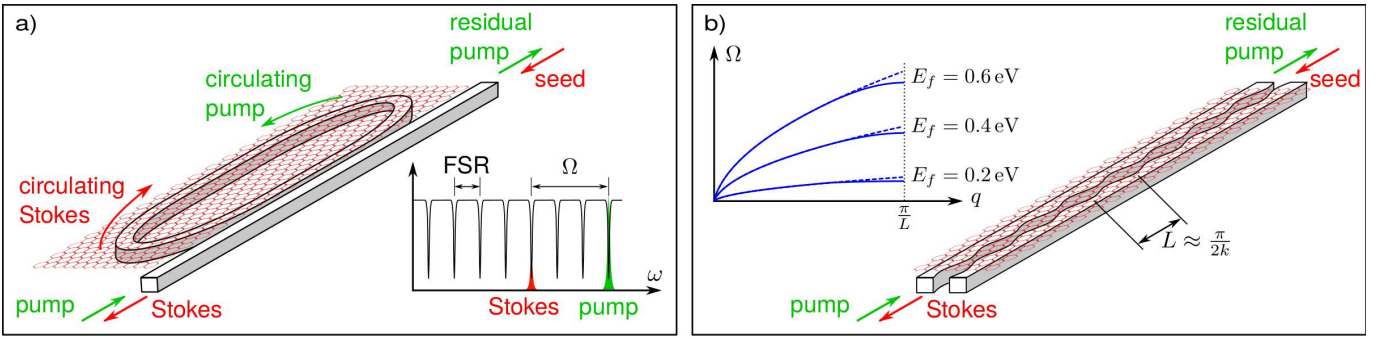}
  \caption{
    Sketches of two possibilities to boost the SPPS response via polaritonic 
    dispersion engineering:
    \\
    (a)
    A ring-topology for the SPPS-active element side-coupled to a bus waveguide
    increases the gain dramatically, but requires the Stokes shift $\Omega$ to
    align with a multiple of the free spectral range (FSR) of the ring, 
    significantly restricting any tunablility.
    \\
    (b) 
    A corrugation at $1/4$ the pump wavelength in the waveguide introduces a
    ``slow-light'' regime with corresponding Purcell enhancement for the
    polariton mode (solid blue lines compared to dashed for the uncorrugated 
    waveguide).
    As the enhancement appears at a fixed wave number, the \emph{polaritonic}
    Purcell enhancement is controlled by the \emph{pump} frequency and can be
    achieved virtually irrespective of the actual Stokes shift 
    $\Omega$, which remains tunable \eg via the Fermi level.
  }
  \label{fig:enhancement}
\end{figure*}

The example system of Fig.~\ref{fig:example_system} was selected for its 
simplicity, maintaining conservative numbers for the aspect ratios and feature 
and gap sizes. 
Besides the technical problem of transferring a graphene-sandwich, the main
experimental challenge is the linear loss introduced by the graphene.
For a given photon energy (\eg $0.8\,\text{eV}$ as chosen in our example) and
variable Fermi level, this introduces two quite distinct regimes:
If the Fermi level is below half the photon energy (here: $0.4\,\text{eV}$),
graphene is extremely lossy due to inter-band transitions, leading to high 
linear waveguide loss (in our example reaching power loss values up to
$10000\,\text{dB/cm}$).
For Fermi levels above half the photon energy, however, the loss drops by
several orders of magnitude (we calculate values as low as $1\,\text{dB/cm}$ 
at $60\,\text{K}$ and just above the inter-band threshold).
Since the linear optical loss limits the useful waveguide length, a natural
figure of merit is the ratio between gain and loss:
% 148 words
\begin{align}
  F = G / \kappa_a,
\end{align}
which takes units of $\text{W}^{-1}$ and whose inverse $1/F$ is the pump power
level necessary to achieve net gain (the effect can remain detectable at 
considerably lower pump levels \eg by pump modulation and lock-in amplification
of the output Stokes).
We plot this in Fig.~\ref{fig:example_system}c.
For most Fermi levels, we find values around $0.1 \ldots 1\,\text{W}^{-1}$, 
which is enhanced up to $100\,\text{W}^{-1}$ due to a ponderomotive resonance 
at the inter-band threshold.
This means that peak pump powers from $10\,$W down to $10\,\text{mW}$ would 
provide net-gain (the latter at liquid nitrogen temperatures and for waveguide 
lengths in the low millimeter range).
This is very feasible with state-of-the-art silicon photonics techniques and
sub-ns pulsed light sources would prevent damage due to excessive dissipation
in the graphene and free carrier absorption in silicon.
  We note that for pump powers on the higher side, pump dissipation in the
  graphene causes a relevant difference between the temperatures of the 
  electron gas and the phonon gas.
  This leads to reduced SPPS-gain and increased \emph{optical} loss while
  leaving the polaritonic quality factor for the most part unchanged.
  The exact threshold for this effect depends strongly on optical pump power, 
  (lattice) temperature, Fermi level and waveguide design (see the Supplementary
  Note 2 for further discussion).
  For our configuration ``1'' at $60\,\text{K}$ with $E_F\geq 0.4\,\text{eV}$
  we estimate a temperature difference of $10\,\text{K}$ at $300\,\text{mW}$ 
  pump power.
  To illustrate the effect, we show in Fig.~\ref{fig:example_system} also
  the impact of a very strong imbalance of $100\,\text{K}$ expected for 
  $3\,\text{W}$ pump at $60\,\text{K}$.
  We note that this power level is already close to the destruction threshold 
  of a silicon waveguide even with short picosecond pulses, so we present an
  extreme case.
  Finally, we find that the overall performance declines above 
  $E_F \approx 0.5\,\text{eV}$ despite in continuously increasing plasmonic 
  quality factor.
  This is due to the resonant ponderomotive interaction and means that doping 
  beyond the currently feasible levels would not provide any further benefit.
% 116 words

We will now present some guidelines for the design of a more sophisticated 
geometry with superior performance over our simple slot-waveguide example.
Firstly, we point out that the scaling between the polariton wavenumber $q$ and 
the quality factor $\Omega / (2 v_b \kappa_b)$ is counterintuitive to people
from the opto-mechanics community.
In SBS, acoustic loss grows super-linearly as $q$ is increased, leading to a 
decrease in the quality factor.
As a result, SBS-active elements are ideally designed to have a low optical
mode index and high acoustic mode index.
Curiously, we find the opposite effect in SPPS: the polaritonic quality factor
\emph{increases} as the wave number and frequency are increased. 
We believe this to be a direct result of decreasing polaritonic mode 
confinement.
As a result, high-mode-index optical waveguide are more viable and
especially the use of plasmonic waveguides based on noble metals appears 
quite interesting.
Even the small propagation lengths are not necessarily a problem, because
of the high gain and the inherent high optical loss of materials such as 
graphene in the inter-band regime.
Furthermore, we point out that any form of field enhancement will improve the
relative gain figure $F$.
While it is true that placing the polaritonic sheet in higher field 
enhancements inevitably increases the linear loss due to inter-band 
transitions, the SPPS process (like SBS and Raman scattering) is effectively a
third-order nonlinearity and thus is bound to over-compensate the increase in 
loss.
% 220 words

Secondly, we find that dielectric loading of the polaritonic mode is quite 
beneficial.
This is the reason why a graphene sheet positioned above a slot waveguide 
provides gains that are comparable to that of a narrow ribbon squeezed inside
the gap, even though it is subject to considerably lower intensity enhancement
and overall poorer mode overlap.
It is not clear how screening of the polaritonic mode (\eg due to the use of
gold-plasmonic waveguides) affects the overall gain.
% 77 words

Finally, we anticipate that the gain can be dramatically boosted by dispersion 
engineering of both the optical and the polaritonic mode.
One example for such a system is to shape the SPPS-active element into a ring
that is side-coupled to a bus waveguide (Fig.~\ref{fig:enhancement}a).
This topology has been studied thoroughly in the context of SBS and can 
dramatically increase the overall gain, even allowing for spontaneous 
oscillation (lasing)~\cite{Mirnaziry:2017}.
However, it requires careful choice of the coupling parameter and close matching
between the Stokes shift and the ring's free spectral range.
As a result, it is a well understood and easily implementable concept, which
however negates the opportunities offered by tuning the Fermi level.
A second possibility is to corrugate the waveguide to introduce a band edge 
with associated slow-light regime.
This is known to enhance SBS in slow-light fibers~\cite{Merklein:2015}.
Alternatively, it is also possible to create a polaritonic band edge
(Fig.~\ref{fig:enhancement}b), reducing 
the polaritonic group velocity and hence mode power $\Power_b$ that appears in 
the denominator of Eq.~\eqref{eqn:gain}.
Despite the resonant nature of such a ``slow-light'' regime, this Purcell 
enhancement would be in fact \emph{broad-band}.
This may seem counter-intuitive at first, but is just a result of the fact that
the band edge is always positioned at the Brillouin zone border, \ie for a 
fixed wave number.
Since $q\approx 2 k$ is effectively fixed by the optical pump wave and 
$\Omega$ follows \eg as a function of the Fermi level, a corrugation with 
a period of $1/4$ of the pump wave length in the waveguide will \emph{always} 
introduce a Purcell enhancement irrespective of the value of $\Omega$.
% 248 words

Beyond the value of SPPS as a tool for the fundamental science of atomically 
thin materials, we also anticipate potential practical applications once 
geometries with optimized gain have been developed.
The first, obvious possibility is to adapt some of the current applications of 
SBS such as integrated light sources (similar to Raman-lasers), optical signal 
filtering and especially sensing, which is one of the areas where SBS is 
currently commercially applied.
Like its acoustic counterpart, the polaritonic dispersion relation is highly
sensitive to variations in the surrounding material.
While the current SBS sensors mainly detect variations of the cladding's 
acoustic impedance and static strain fields,
similar SPPS sensors involving a sheet of 2d material would be highly sensitive 
to very thin adsorbed layers either through a change of permittivity, a
reduction in the carrier mobility or shifts in the Fermi 
level~\cite{Xiao:2016}.
Furthermore, the principles of Brillouin optical correlation domain analysis 
(BOCDA) could be adapted to pinpoint the perturbation along an SPPS sensor with 
sub-mm resolution~\cite{Zafiri:2018}.
% 164 words

Finally, we would like to emphasize another aspect that makes SPPS interesting 
for practical applications: the amplification and detection of signals in a 
narrow frequency band that can be selected anywhere in the THz- to mid-IR band.
Like SBS, SPPS is a self-amplifying process that increases the amplitude of both
the low-frequency excitation (sound in the case of SBS, polaritons in SPPS) as
well as the optical Stokes signal exponentially along the waveguide.
This means that a weak THz signal injected into a graphene sheet at the 
back of an SPPS-active waveguide could be detected in the optical domain as 
the Stokes signal.
Additionally, the amplified THz-signal could be picked off at the front of the 
waveguide.
Both the detection and amplification would be restricted to the narrow 
frequency window given by the plasmon frequency and quality factor at the 
given wave number and could be tuned within a wide range via the Fermi level.
Such a tunable narrow-band detector in conjunction with a wide-band source (\eg
a thermal emitter) would have the same versatility as a narrow-band source in
conjunction with a broad-band detector, and could be of considerable value to 
spectroscopy in the far-IR and THz regime.
% 199 words

% Summary
In summary, we described the optical coupling to (\eg plasmonic) polaritons in 
2d and vdW materials through a new physical process that we call stimulated plasmon 
polariton scattering.
We outlined the theoretical framework, and based on experimentally verified 
material parameters we showed that the process can be observed in an 
experimentally straight-forward system at moderate temperatures despite significant 
optical loss and that even net gain for a Stokes signal can be achieved
with a continuous-wave pump in this example at the appropriate Fermi level. 
% 71 words

\section*{Acknowledgments}
C.~W. acknowledges funding from a MULTIPLY fellowship under the Marie
Sk\l{}odowska-Curie COFUND Action (grant agreement No. 713694).
The Center for Nano Optics is financially supported by the University of
Southern Denmark (SDU 2020 funding).
N.~A.~M. is a VILLUM Investigator supported by VILLUM Fonden (grant No. 16498).
The Center for Nanostructured Graphene is sponsored by the Danish National 
Research Foundation (Project No. DNRF103). 
We are deeply indebted to P.A.D.~Gon\c{c}alvez, Joel Cox, and especially 
C.~Tserkezis for valuable comments on the manuscript.
% 76 words

\section*{Supplementary Information}
The supplementary notes can be requested from the authors via email.

\end{document}